# Post-mortem analysis of tungsten plasma facing components in tokamaks: Raman microscopy measurements on compact, porous oxide and nitride films and nanoparticles.

C. Pardanaud[a*], D. Dellasega[b,c], M. Passoni[b,c], C. Martin[a], P. Roubin[a], Y. Addab[a], C. Arnas[a], L. Couëdel[a, d], M. Minissale[a], E. Salomon[a], G. Giacometti[a], A. Merlen[e], E. Bernard[f], R. Mateus[g], E. Alves[g], Z. Siketic[h], I. Bogdanovic Radovic[h], A. Hakola[i], and EUROfusion WP PFC contributors[j]

[a]*Aix Marseille Univ, CNRS, PIIM, Marseille, France*
[b]*Dipartimento di Energia, Politecnico di Milano, Milano, Italy*
[c]*Istituto di Fisica del Plasma "P.Caldirola", Consiglio Nazionale delle Ricerche, Milano, Italy*
[d]*Phys. and Eng. Phys. department, University of Saskatchewan, Saskatoon SK S7N 5E2, Canada*
[e]*Aix Marseille Univ, Université de Toulon, CNRS, IM2NP, Marseille, France*
[f]*CEA, IRFM, F-13108 Saint Paul lez Durance, France*
[g]*Instituto Superior Técnico, Universidade de Lisboa, Bobadela, Portugal*
[h]*Rudjer Boskovic Institute, P. O. Box 180, 10002 Zagreb, Croatia*
[i]*VTT, P. O. Box 1000, 02044 VTT, Finland*
[j]*See the author list in "S. Brezinsek et al 2017 Nucl. Fusion 57 116041"*

**Abstract**

Raman microscopy is one of the methods that could be used for future post-mortem analyses of samples extracted from ITER plasma facing. This study shows that this technique is useful for studying tungsten-based materials containing impurities including oxides and nitrides. Here, we apply pulsed laser deposition and DC argon glow discharges to produce tungsten-containing synthetic films (compact, porous) and nanoparticles and investigate the influence of their morphology on the measured Raman spectra. The amounts of oxygen and/or nitrogen in the films are also investigated. Comparative data are obtained by X-ray Photoelectrons Spectroscopy, Atomic Force Microscopy, Electron Microscopies (Scanning and Transmission), Energy Dispersive X-ray spectroscopy, Time-of-Flight Elastic Recoil Detection Analysis. The power density of the laser beam used to perform Raman microscopy is varied by up to 4 orders of magnitude (0.01-20 mW/$\mu m^2$) to investigate thermal stability of films and nanoparticles. As a first result, we give evidence that Raman microscopy is sensitive enough to detect surface native oxides. Secondly, more tungsten oxides are detected in porous materials and nanoparticles than in compact films, and the intensities of the Raman band correlate to their oxygen content. Thirdly, thermal stability of these films (i.e. structural and chemical modification under laser heating) is poor when compact films contain a sufficiently large amount of nitrogen. This finding suggests that nitrogen can be substituted by oxygen during Raman laser induced heating occurring in ambient air. Finally, our methodology can be used to rapidly characterize morphology and chemistry of the samples analyzed, and also to create oxides at the micrometer scale.

1. **Introduction**

Tungsten is one of the elements that will be used in the fabrication of plasma facing components (PFC) for the ITER tokamak. It has been chosen due to its good thermomechanical properties and low affinity with hydrogen isotopes in order to face the high heat due to nuclear fusion reactions between hydrogen isotope nuclei [1, 2]. Heat fluxes, in the range 10 to 20 MW/m$^2$ in steady state operations and even higher (in the GW/m$^2$) for edge localized modes or other transients will have to be radiated off [3, 4], mainly on the divertor [5]. However, due to plasma wall interactions [6], PFC characteristics will be modified during the lifetime of the machine [7] due to the many simultaneous processes (erosion, morphological changes, melting, defect formation…), involving high heat fluxes, impinging ions and neutrals [8] formed by hydrogen isotopes (fuel of the reaction), helium (ashes of the reaction), neutron bombardment… and will modify the PFCs. The presence of impurities such as oxygen or nitrogen, and eroded species that can migrate or redeposit [9, 10] should also be taken into account. As a result, PFCs will react chemically (for example forming oxide layers in specific conditions [11, 12], with seeding gas [13, 14]), melting being a possibility that can enhance those chemical effects. As a consequence, the safety conditions of ITER may be compromised, mainly due to the formation of dusts [15-17] and retention of radioactive tritium used to fuel the plasma [18-20]. The prediction of how these modifications will alter long term operations is thus a keystone in the field. To this end, two complementary strategies are generally used. The first one involves testing scenarios in existing tokamaks, e.g., in the JET tokamak with ITER like wall (ILW) configuration [21] or in the WEST tokamak [22, 23], and then using scaling laws to predict what could happen under ITER conditions. The second one involves testing for PFC and/or laboratory materials for one specific or several synergetic effects together, in facilities [24, 25] allowing controlled parameters (surface temperature, morphology, composition, ion flux and energy…) [26-29]. In this article, synthetized materials are produced to mimic deposits or modified surfaces [6].

Post-mortem analyses are then used for all the samples to obtain a before/after comparison. Traditionally used techniques include ion beam analyses [30] and thermal desorption spectroscopy that give the concentrations and depth distributions of various elements, electron microscopy, that gives information on the morphology/porosity and X-ray diffraction (XRD) that reveals the structure of and nature of traps in the material [31, 32].

In addition, we use Raman microscopy. This technique is based on an inelastic light scattering process and is highly sensitive to chemical composition, structure and morphology. Depending on the optical constants of the probed material, Raman microscopy can be sensitive to the near surface (few tens of nanometers in the case of metals) with a micrometer lateral resolution, that allows monitoring spatial inhomogeneities [11, 12]. Raman microscopy method gives information about atomic vibrations, and is thus able to distinguish chemical bonds in the solids and to reveal how hydrogen isotopes and impurities are bonded, either to metal atoms or to each other's [33, 34]. It is particularly sensitive to detect tungsten oxides and the modifications of their characteristic structure and composition after He or D implantation [35-37], as well as to detect the presence of H in oxides [38]. It is also sensitive to the crystalline structure and the presence of defects [39].

In this work, we focus on both the detection of impurities in tungsten by means of Raman microscopy and the influence of material surface roughness on the measured spectra. We use a large variety of laboratory samples: pulsed laser deposition (PLD) grown W films with different morphologies (porous or compact) and different chemical compositions (with different O and N contents), W nanoparticles and a W bulk ITER-grade sample and thick $WO_3$ layers. This work is a continuation of our previous work [40] demonstrating that Raman microscopy gives a reliable detection of oxides for Be based samples [11, 12] and for W based samples. It is therefore a well-suited technique for post-mortem analyses of ILW PFCs. In addition, in this study, by varying the power of the incident laser by up to 4 orders of magnitude, we can study the fingerprint of laser irradiation on the Raman spectra. It allows us quickly retrieve information about the morphologies of O and N containing W films. This is due to structural changes that laser heating induces when thermal dissipation is not efficient, which is the case of porous samples.

## 2. Methods
### 2.1 Sample preparation

For this study, we synthetized and analyzed 11 W based samples, containing various amounts of O and N. We compared them to a polycrystalline tungsten sample provided by *A.L.M.T. Corp* (Japan) with a mirror-like polishing and a set of $WO_3$ tungsten oxide layers obtained by thermal oxidation of W under $O_2$ pressure (see [35] for details) with thicknesses varying from 20 to 200 nm. From the batch of 11 samples, 1 was composed of agglomerated nanoparticles, deposited by magnetron DC sputtering (see [41] for details), while the 10 others were produced by Pulsed Laser Deposition (PLD). Varying the experimental conditions allowed changing their chemical composition (amount of O in the bulk varied from 0 to 60 atomic % and/or the amount of N from 0 to 20 atomic %), and their morphology (compact or porous with a cauliflower like structure). The PLD samples were produced as thin films on metallic substrates (stainless steel or Mo) using a 532 nm laser in the nanosecond regime (5-7 nm, 10 Hz, pulse energy 700 mJ), focused on a W target (purity 99.99%) with an energy density of 15 J/cm$^2$. The species ablated from the target expanded in the vacuum chamber (base pressure of $10^{-3}$ Pa) and ended up on the substrate, located 60 mm away from the target. All the films were deposited at room temperature. The depositions were performed in different background atmospheres and pressures (Ar, He and $N_2$ and with pressure in the range 2.5-70 Pa). The use of different process gas results in different film morphologies: compact amorphous (c-) or porous (p-) and compositions. Further details are reported in [42, 43]. Thickness of these samples was ≈ 1 µm. The synthesis conditions and bulk compositions are reported for the samples studied in table 1.

## 2.2 Sample characterization

The morphology of the 10 PLD samples has been assessed by a ZEISS Supra 40 scanning electron microscope (SEM) using an accelerating voltage of 5- 7 kV, as will be presented in more details in section 3. Nanoparticles have been studied using transmission electron microscopy (TEM). A Cs corrected FEI 80–300 device is used to acquire high resolution electron micrographs (HR TEM) [41]. Bulk elemental composition of the 10 PLD films has been obtained by Time-of-Flight Elastic Recoil Detection Analysis (TOF ERDA) and/or Energy Dispersive X-ray spectroscopy (EDX). The EDX system used is an Oxford Instruments Si(Li) detector in combination with the SEM microscope. The TOF-ERDA measurements were performed using a 6 MV Tandem Van de Graaff accelerator located at the Ruđer Bošković Institute, Zagreb. For the analysis 20 MeV $^{127}I^{6+}$ ions at a 20° incidence angle toward the sample surface were used, allowing to measure down to 300 nm in the bulk. The TOF-ERDA spectrometer was positioned at an angle of 37.5° toward the beam direction. More details about the experimental setup can be found in ([44, 45]). Note that the samples $c$WO(25,2) and $p$WO(60,2.5) were previously analyzed in [46]. Atomic force microscopy (AFM) was used to study the roughness of the samples. Here, the tapping mode of an NTMDT solver was used, the tip radius was ≈ 10 nm, and the vertical and horizontal resolutions were ≈ 1 and ≈ 10 nm, respectively. X-ray Photoelectrons Spectroscopy (XPS) analyses were performed at the PIIM laboratory. The experimental setup is composed of a high-intensity twin (Mg/Al) anode X-ray Source (provided by PREVAC) emitting at 1253.6 and 1486.6 eV respectively. The analyzed samples were placed in a movable sample holder inside a mu-metal ultra-high vacuum chamber (base pressure $1\times10^{-10}$ mbar). Photoelectrons were detected via a high-resolution electron energy analyzer (Scienta Omicron R3000) composed of a high transmission electron lens with an acceptance angle of 30 degrees and a 40 mm MCP detector monitored by a FireWire CCD-Camera. The resolution of the XPS spectrometer, determined from the full-width at half maximum of the Ag 3d core levels of a clean Ag single crystal, was 0.9 eV.

Raman spectra were recorded in the back-scattering geometry using a Horiba-Jobin-Yvon HR LabRAM HR800 Raman microspectrometer, using one laser wavelength $\lambda_L$ = 514.5 nm. The used objectives had a ×100 magnification, with a numerical aperture (N.A.) of 0.9 and ×50 magnification with a numerical aperture of 0.5. The radius of the laser spot with N.A. = 0.9 is R ≈ 0.6 × $\lambda_L$ / N.A ≈ 0.34 µm. To study possible heating effect, the laser power density was tuned from 0.01 to ≈ 17 mW µm$^{-2}$, using the mapping mode for low power density in order to avoid any oxidation due to the probe beam as measurements were performed in air. The used grating (600 lines/mm) gave a spectral resolution of ≈ 1 cm$^{-1}$.

**2.3 Sensitivity and capability of Raman spectroscopy for studying oxides and nitrides**

As there are only acoustic phonon branches in tungsten, and due to the spectroscopic configuration used in this study, no Raman bands are supposed to be detected, meaning this spectroscopic technique is not adequate to probe pure W. However, oxides, even if not wanted, will be present in tokamaks [12, 47]. Then, Raman spectroscopy signal is expected to be very rich for oxides and bronzes (i.e. hydrogen in oxides) as the numerous structures listed in the literature lead to optical phonon branches. In figure 1, some typical oxide spectra are shown as well as the phonon density of states (PDOS) of pure W, which can be seen due to relaxing quantum selection rules when defects are introduced [48]. Whatever the O/W stoichiometry, the spectra are mainly composed of bands related to bending and stretching modes of W-O-W bonds, in addition to a band related to the vicinity of the surface [49-51]. To better understand the spectra, we need to go deeper into the details of the tungsten oxide structure. The building blocks of tungsten oxide are $WO_6$ octahedra: 1 W and 4 surrounding O form a basal plane, and 2 out of plane O, above and below that plane, form the octahedra. This local arrangement governs the main trends observed in the Raman spectra, especially the bands lying at 270, 714 and 805 $cm^{-1}$. However, these octahedra are corner- or edge-sharing to form the numerous tungsten oxide polymorphs observed in the nature and this can affect the Raman spectra as symmetry, stoichiometry and interatomic distances change. Depending on the tilting angle and rotation direction of the octahedra relative to the others [52], several $WO_3$ phases can exist. For example, for bulk $WO_3$, triclinic δ-$WO_3$ is the existing phase at low temperatures. From 17°C up to 330°C, the monoclinic phase γ-$WO_3$ dominates, then the orthorhombic β-$WO_3$ phase up to 740°C where the tetragonal α-$WO_3$ starts to form. The transition temperatures can decrease when nanostructures appear [39, 49] and/or when the stoichiometry changes slightly from that of a trioxide [53]. Magneli phases with stoichiometries close to trioxide, from 2.625 to 2.92, have been reported earlier [54]. The aim of figure 1 is thus to give a brief overview of the main spectroscopic features and differences that characterize tungsten oxides. The vertical dashed lines give the band positions (270, 714 and 805 $cm^{-1}$) of the three main Raman active modes of bulk monoclinic $WO_3$. Subtle changes in the spectra (band position mainly) can occur depending on the $WO_3$ phase but they are considered negligible at this step, compared to the differences that other kinds of oxides introduce in the spectra, as can be seen with a quick view in this figure. Introducing defects that diminish the crystallite size at which the octahedral periodic pattern is perfectly repeated results in broadening the peaks and increasing the intensity levels of the low frequency bands centered at 270 $cm^{-1}$. This is what can be seen on the spectrum labeled d-$WO_3$, d standing for defective. Details of this sample synthesis can be found in [55]. For nano trioxide (labelled n-$WO_3$, with a crystallite size of 4 nm only, *[39]*), broadening further increases, resulting in the merging of the 714 and 805 $cm^{-1}$ bands, and the intensity of the low wavenumber bands diminishes. For samples labelled $WO_x$ [56], where oxygen content is low compared to that of a trioxide, the stretching mode bands are distinguishable but are overlapping and shifted. The 275 $cm^{-1}$ band, superimposed to a broad weak band, is sharp and close to the value recorded for monoclinic $WO_3$, with no real explanation up to now. For d-$WO_3$, a-$WO_3$ and $WO_x$, a band close to 950-960 $cm^{-1}$ becomes visible. It is interpreted as W=O bond, due to electronic reorganization when neighbors are missing at the surface or at the limits of crystallites. A sub trioxide labelled $WO_{3-x}$ (possibly $WO_{2.9}$ according to [57] but it could also be $WO_{2.72}$) and $WO_2$ (formed by heating the d-$WO_3$ up to 800°C in vacuum), are also shown, for comparison. The literature on tungsten nitride is not as rich as the one on oxides, may be due to much fewer applications for nitrides. However, cubic $W_2N$, which is supposed to be the most stable phase under normal conditions, is not considered Raman active. When the stoichiometry is not preserved, the W phonon density of states has been observed to rise close to 200 $cm^{-1}$, the corresponding bands being roughly ten times more intense than other bands involving W-N bonds and lying at 471 $cm^{-1}$ and in the range 700-800 $cm^{-1}$ [48]. Then, for an unknown sample possibly containing both N and O, if bands are observed close to 800 $cm^{-1}$, they can in principle involve both W-N and W-O bonds. Depending on the relative amounts of N and O, it can be hard to disentangle the chemical origin of this band. Fortunately, the shape and intensity of the bands below 300 $cm^{-1}$ behaves differently for oxides and nitrides. For example, the W PDOS is very intense compared to the rest of the bands in the presence of nitrides, which is not the case for oxides.

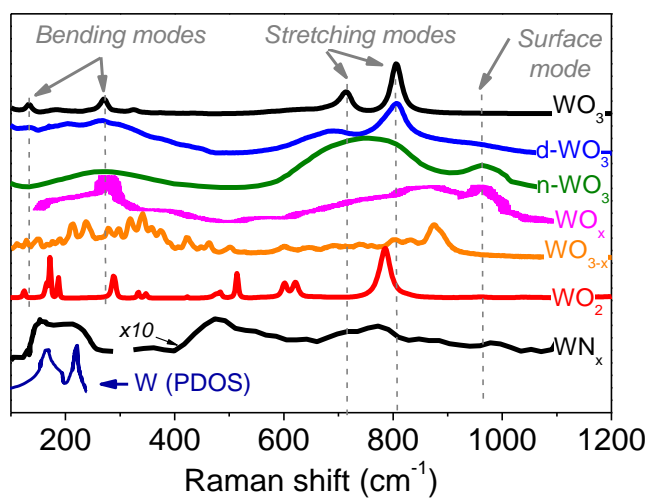

*Figure 1. Raman spectra of usual tungsten oxides and nitrides and the W phonon density of states (PDOS) of pure tungsten. Here, d and n stand for defective and nano, respectively. n-$WO_3$ and $WO_x$ spectra are from [39] and [56], respectively. $WN_x$ spectrum is from Ref. [48]. Above 300 $cm^{-1}$, the intensity has been multiplied by 10 to make the bands visible. W PDOS is from Ref. [58].*

## 3. Results

The content of impurities (O and N) is one of the main parameters varied in this study (see table 1). For O, it ranges from trace levels when only surface native oxides are formed (the case of *c*W sample), up to 75 at. % in the bulk for $WO_3$ reference layers, with intermediate concentrations varying from 25 to 60 at. % for the PLD samples. Another parameter to be altered was the morphology of the samples. Raman band intensities have also to be considered in order to obtain information on the O and the N contents. Note that this aspect is in general not addressed in the literature as optical properties and electromagnetic wave propagation in the materials need to be precisely known [59], and roughness can play an important role which is difficult to take into account. We present morphology and chemical composition analysis in section 3.1, information retrieved from absolute intensities of oxide related bands in section 3.2 and the effect of the laser beam power in Raman excitations in section 3.3.

| Sample | Synthesis method | Elemental composition |
|---|---|---|
| *c*W bulk | From A.L.M.T | O 0%, N 0% (purity: 99.9999 wt%) |
| *c*WO(25,2) | PLD (He, P= 70 Pa) | O 25%, N 2%, H 2% (TOF ERDA) |
| *c*WN(0,20) | PLD (N, P= 2.5 Pa) | O 0%, N 20% (EDX) |
| *c*WON(37,16) | PLD (He 90% +N 10%, P= 55 Pa) | O 37%, N 16%, H 8% (TOF ERDA) |
| *c*WON(20,35) | PLD (N, P= 20 Pa) | O 20%, N 35% (EDX) |
| *p*WO(60,2.5) | PLD (Ar, P= 50 Pa) | O 60%, N 2.5%, H 15% (TOF ERDA) |
| *p*WO(52,4) | PLD (Ar 97% +D 3%, P= 50 Pa) | O 52%, N 4%, H 19% (TOF ERDA) |
| *p*WON(53,12) | PLD (Ar 97% + N 3%, P= 50 Pa) | O 53%, N 12%, H 19% (TOF ERDA) |
| *p*WON(47,9) | PLD (Ar 94% +N 3% + D 3%, P= 50 Pa) | O 47%, N 9%, H 24% (TOF ERDA) |
| *p*WON(20,35) | PLD (N, P= 40 Pa) | O 20%, N 35% (EDX) |
| NP-W | DC argon glow discharge | - |
| $WO_3$ on W (20-200 nm) | Thermal oxidation at 400°C [35] ($O_2$ partial pressure: from 0.7 to 79 kPa) | - |

*Table 1. Sample description. Names starting with c (p) correspond to a compact (porous) morphology, as observed by electron microscopy. The first number in parenthesis is the O content and the second is the N content, expressed in atomic percentage.*

### 3.1 Morphology and chemical composition

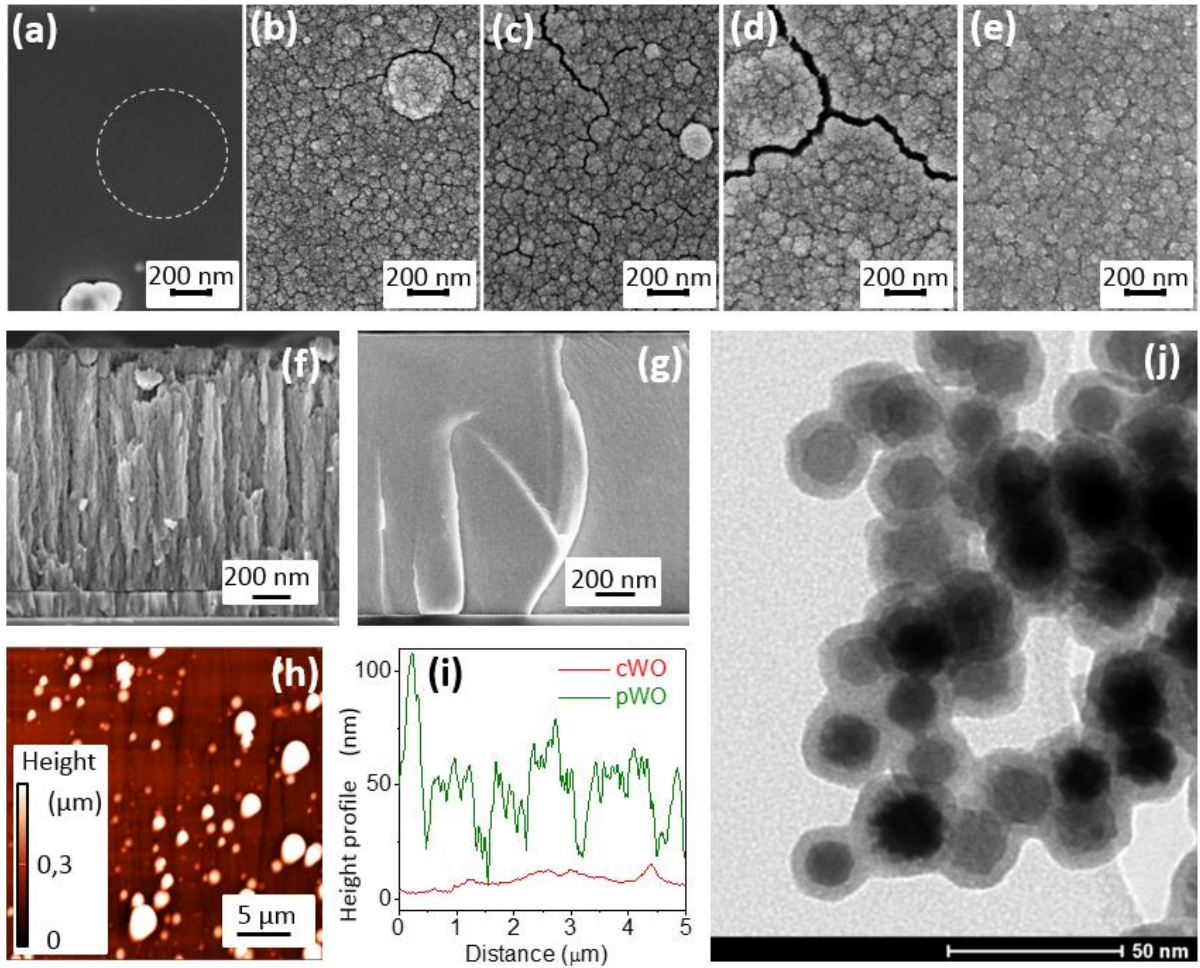

**Figure 2.** *Morphology of the PLD films and the nanoparticles. (a-e) PLD films top view, the dashed white circle represents the size of the laser beam used in the Raman experiments when using an objective with N.A.=0.9. From a to e: cWO(25,2), pWO(60,2.5), pWO(52,4), pWON(53,12) and pWON(47,9). (f, g) Typical SEM cross section view of PLD films, porous (f) and compact (g). (h) Typical AFM top view of porous films. (i) Typical AFM profiles of porous and compact films. (j) Nanoparticles.*

Figure 2 shows SEM and AFM images of the PLD films and TEM images of the nanoparticles (NP). Top views of PLD films, cWO(25,2), pWO(60,2.5), pWO(52,4), pWON(53,12) and pWON(47,9) can be seen in Fig-1 a to e, respectively. Typical porous and compact film SEM cross sections (obtained after cleaving the samples) are displayed in figure 1-f and g, respectively. A 30×30 µm$^2$ AFM image of a typical porous film is displayed in Fig. 1-h and typical height profiles of porous and compact films are given in Fig. 1-i. A TEM image of nanoparticles is shown in Fig. 1-j, which reveals a core-shell structure. From figure 1-a to e, one can observe that the morphology of the *c*WO film is different from that of porous films: the *p*WO(60,2.5), *p*WO(52,4), *p*WON(53,12) and *p*WON(47,9) films show a cauliflower-like feature whereas the surface of the *c*WO(25,2) film is flat and smooth. This morphology is directly related to growth mechanisms [60] governed by the expansion dynamics of the plasma plume during deposition, as discussed elsewhere [42, 43]. Some cracks are present on the porous films. Cross sections are shown in figures 1-f and g, showing the 1 µm thickness of the films. Columnar structures are better visible, at the origin of the porous label of the *p*WO(60,2.5), *p*WO(52,4), *p*WON(53,12) and *p*WON(47,9) films. AFM measurements confirm the rough morphology of the porous films (≈ 25 nm amplitude obtained on profiles by measuring the peak to peak height) compared to that of the compact samples (≈ 7 nm). Some columns higher than the mean surface by ≈ 200-300 nm can be seen in Fig 1-b (top right part), f and h (white spots are not due to dust). Raman spectra were not acquired on these columns to avoid probing non-representative locations. As reported in [42, 46] both *c*WO and the different *p*WO and *p*WON films exhibit a nanocrystalline structure where the crystallite domains range between 4 and 7 nm. Nanoparticles (Fig 1-j) display a polyhedral form, sometimes nearly spherical, with diameters ≈ 14-20 nm and a core composed of W in the β phase [41], and an oxide shell with a thickness of ≈ 4-5 nm.

XPS spectra of selected W, WN, compact WO, porous WO samples are shown in figure 3-a and b (XPS was not possible for NP as there was not enough material). Their Raman spectra are given in figure 3-c. The XPS spectra recorded in the 4f region are reported in figure 3-a for bulk oxide samples (d-WO$_3$, *p*WO(60,2.5) and *c*W(25,2)) and in figure 3-b for surface oxides (*c*W) and for nitrides (*c*WN(0,20)). According to the 2012 update of the NIST database [61] 4f$_{7/2}$ and 4f$_{5/2}$ lie at 31 eV and 33 eV, respectively for W [62] whereas they lie at around 36 and 37.5 eV for WO$_3$ [62, 63]. An additional shift of roughly 0.5 eV has been reported in the literature [53], possibly due to nano structuration and phase changes; we do not discuss in this publication. The XPS spectrum of d-WO$_3$, extracted from [37], is composed of two peaks related to WO$_3$ but slightly shifted (it can be due to a variation of the work function between the different samples or setup calibration). The XPS spectra of *p*WO samples display mainly two peaks at 36 and 37.5 eV, originating from oxide formation but are broader and more overlapped compared to those of d-WO$_3$. A weak peak observed at 30.8 eV for *p*WO is due to bulk W. Such measurement is consistent with the stoichiometry measured by TOF-ERDA. The XPS spectra of *c*WO samples shows a combination of oxide and W components, meaning coexistence of an oxide phase and a W phase, in different proportion than for *p*WO samples, still consistent with TOF-ERDA measurements. For samples supposed to have no oxide in the bulk, XPS spectra are shown in figure 3-b. The XPS spectrum of the *c*W sample is composed of 2 intense peaks lying at 31 and 33 eV plus two weaker peaks at 35 and 37.5 eV. These indicate a low amount of oxide at the surface of the film (here, XPS roughly probes a depth of 1.5 nm). The XPS spectrum of the *c*WN(0,20) sample is composed of 4 broad overlapping peaks, situated approximately where WO and W signals are expected. Some studies suggest that WN signals are in the range 31.5-31.7 eV [64] whereas others mention 32.6 and 34.7 eV [65]. One must note that the most intense peaks in figure 3.b are the 35.7 and 33.7 eV, close to where WN should fall according to [65]. In any case, the WN signal overlaps with W and WO signals, and the shown spectrum corresponds to a mixture of W, WN and WO. It reveals that the *c*WN film is oxidized within the probed XPS depth range, in a much larger extent than the *c*W film. Note that the oxidation levels can also affect the global shape of the XPS spectra: for example W$^{4+}$ are close to 4f$_{5/2}$ and 4f$_{7/2}$ related to W, and W$^{5+}$ are in between 4f$_{5/2}$ and 4f$_{7/2}$ related to WO$_3$ [37], and fitting with the resolution used will not allow quantitative results.

In figure 3-c, the Raman spectrum recorded for the *c*W sample display bands lie in the bending (300 cm$^{-1}$) and stretching (700-800 cm$^{-1}$) spectral regions, confirming the presence of oxide. There is also an extra surface band (950 cm$^{-1}$) which is very intense, proving the presence of defective W=O bonds. In W crystals, the penetration depth of visible light is supposed to be only a few tens of nm due to the metallic nature of W [66]. Therefore, the origin of this band is from W-O bonds close to the surface, formed due to the exposure to air. Comparison with other materials (*p*WO(60,2.5), *c*WO(25,2), *c*WN(0,20), NP and defective WO$_3$) is done in this figure. Both *p*WO and NP display an intense 950 cm$^{-1}$ band, showing the existence of a large amount of surface area in these samples, in agreement with what is seen by electron microscopy in figure 2 for *p*WO (Fig. 2 f) and NP (Fig. 2 j) samples. For the *c*W sample, the 950 cm$^{-1}$ band is observed as well, proving that a large surface is present too. For cW, extra well-defined bands are observed close to 400 and 650 cm$^{-1}$.

According to ultra-high vacuum experiments, vibrational modes of O adsorbed on top of W(110) for high surface coverage (750 ML) [67] lie at 428 and 657 cm$^{-1}$. These observations suggest that two contributions leading to WO bonds exist for the *c*W sample: O adsorbed on the W surface and WO bonds organized as small $WO_3$ crystals, leading to a large amount of crystallite surface sites. Note that the features appearing in the dashed rectangle of figure 3-c are not attributed to the samples but to experimental conditions taking into account the nature of the used lens, time acquisition and incident angle [68].

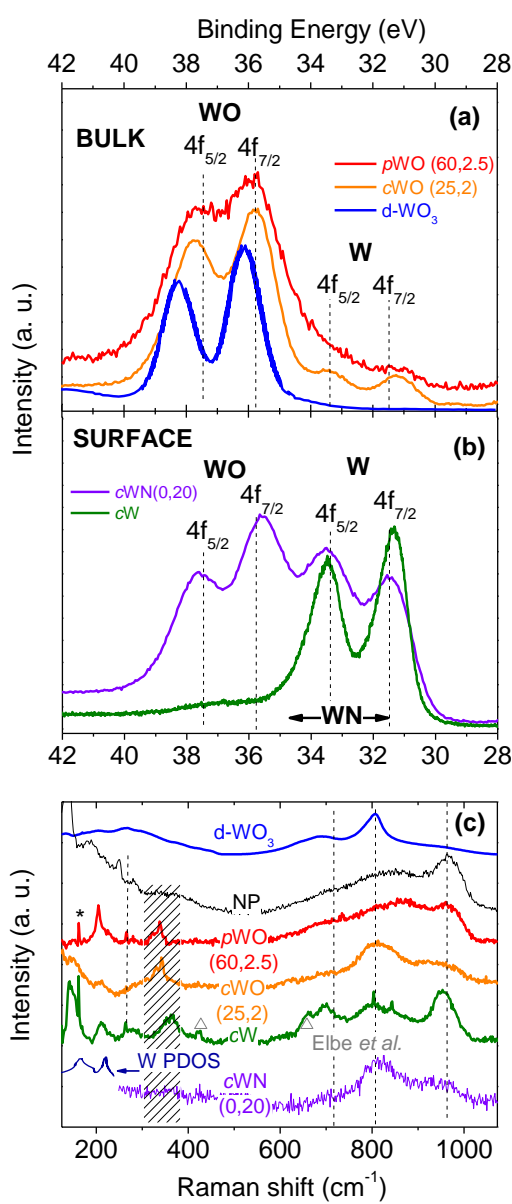

**Figure 3.** *XPS and Raman spectra of d-$WO_3$, NP, pWO, cWO, cW and cWN samples. XPS spectra of (a) pWO(60, 2.5), cWO(25,2) and d-$WO_3$ (the latter from [37]) and (b) cWN(0,20) and cW. (c) Raman spectra. The dashed rectangle of figure 3-c is not attributed to the*

*samples but to the experimental conditions taking into account the nature of the used lens, time acquisition and incident angle [68]*

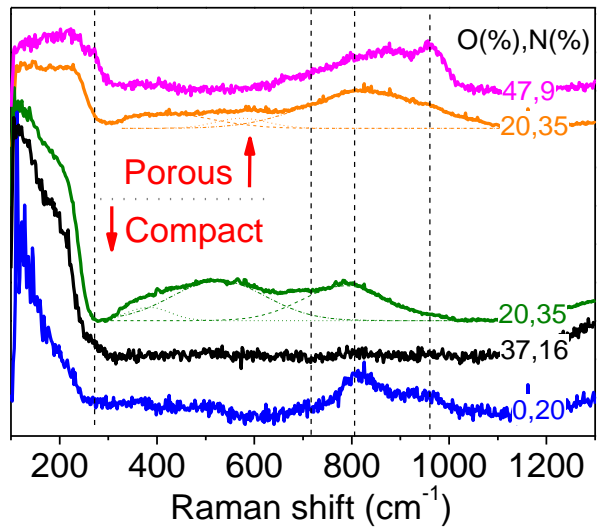

**Figure 4.** *Comparison of the Raman spectra of N containing films.*

Figure 4 shows normalized Raman spectra of different WON samples, either compact or porous. The first observation is that porous and compact samples behave differently and the weight of the W PDOS (below 300 cm$^{-1}$) compared to the O- or N-related bands at higher wavenumbers is much higher for compact samples than for porous samples. A second observation concerns the surface W=O band at 950 cm$^{-1}$. The *p*WON(47,9) spectrum is similar to that of WO$_x$ with a strong contribution of this band: the amount of O is much more important than the amount of N, and in addition nitrides are not supposed to be Raman active. Thus, for low N concentrations, oxygen governs the overall Raman spectral shape in the 300-1000 cm$^{-1}$ spectral region. However, the shape of the bands below 300 cm$^{-1}$ behaves differently than for WO$_x$, which could be due to the presence of N (activating W PDOS), as discussed in section 2.3. When the N/O ratio increases, the behavior changes. For example, *p*WON(20,35) is composed of three broad overlapping gaussians profiles in the range 300 - 1000 cm$^{-1}$. The main one is centered at 833 cm$^{-1}$, whereas the others are centered at 417 and 573 cm$^{-1}$. The first band is interpreted as being partly due to W-N and W-O bonds, because of its position close to both W-N and W-O stretching mode bands, according to figure 1, whereas the two other bands are due to W-N bonds. For the compact cWON(20,35) sample, the 300-1000 cm$^{-1}$ spectral range is also composed of three broad overlapping bands attributed to W-N. But this is not the case for all the N/O ratios. Pure nitrogen doped tungsten, the *c*WN sample, display bands at 813 and 950 cm$^{-1}$, in addition to weaker features around 400 cm$^{-1}$ (broad features between 300 and 550 cm$^{-1}$). The 813 and 950 cm$^{-1}$ bands are interpreted as originating from octahedra and from surface W=O bonds, respectively. The W-N bonds could be related to the very weak signal around 400 cm$^{-1}$. The spectrum from the *c*WON(37,16) sample is totally flat in the same spectral region. In this case, the absence of bands whereas W-N bonds are present in *c*WON(37,16) can be due to some quantum-selection rules related to the structure during the Raman scattering process, whereas the presence of defects (like oxygen and nitrogen, grain boundaries due to small crystallite size,…) can break the related quantum selection rules, leading to the presence of these bands. The lack of signal intensity corresponding to W-O bonds is different here, possibly due to the stability of the W-N bonds that hinder the formation of W-O bonds during the deposition process of the film. This point will be addressed in section 3.2.

To briefly summarize the findings reported above that concern Raman spectroscopy measurements:
- oxides have been detected on a compact W sample, likely due to native oxides (see section 3.2 for sensitivity discussion)
- Compact and porous PLD films exhibit Raman spectra that are distinguishable
- Spectra of porous films and oxides found on nanoparticles are similar, displaying a mode related to surface bonds, lying at 960 cm$^{-1}$
- N content does not lead to strong specific bands but to low frequency modes which are related to the PDOS of W, as found for mixed W:Be deposits [69].
- For some concentrations, oxides are no more observed on W/O/N samples, that could mean the presence of substituting N

**3.2 Information retrieved from Raman absolute band intensities**

While we have discussed information retrieved from the spectral shape, band positions and relative intensities, and the ways to distinguish the presence of N when the sample contains also oxygen, we now discuss the information retrieved from absolute band intensities and its relation to the sample composition. Absolute band intensities are here referred to the height of the band normalized by the height of the band at 520 cm$^{-1}$, H$_{Si}$, of a silicon crystal measured in the same experimental conditions, as this intense band is often used as a reference [40].

In figure 5, the height of the band at 800 cm$^{-1}$, $H_{800}$, is taken on the rough spectra, i.e. without band fitting, and after base line subtraction. The height ratio, $H_{800}/H_{Si}$ is shown in figure 5-a as a function of the oxygen atomic fraction. A general tendency is that the more oxygen the material contains, the more intense the 800 cm$^{-1}$ band, approximately following an exponential law (the dashed line is a guide for the eyes). This trend is well followed for samples that have a similar porous morphology: *p*WO(60,2.5), *p*WO(52,4), *p*WON(47,9), and *p*WON(53,12) and for which nitrogen is not the dominant impurity element. For nanoparticles, the $H_{800}/H_{Si}$ ratio equals 0.12, the highest value recorded in this study. The intersection between the dashed line and this value is at 63% of oxygen. This value is close to the fraction of oxides that can be estimated from TEM images for an average nanoparticle. This indicates that the empirical exponential behavior works qualitatively well for this sample too. When the nitrogen content increases a little, here for the *p*WON(53,12) sample, data are still close to the dashed line, however slightly up-shifted. The shift becomes very large for the high N/O ratio of the *p*WON(20,35) data. Note that this data point could be moved a bit closer to the dashed line if one replaces the x-axis, which is the fraction of oxygen O(%), by O(%)×100/(100-N(%)), which represents the fraction of O bonded to W, excluding the overall tungsten content the ones bonded to N. This transformation is shown in figure 5-b. By doing that, one assumes that W-N bonds do not participate significantly to the intensity of the band close to 800 cm$^{-1}$. Another hypothesis is that O and N are not bonded together but only to W. If this was not the case, it would shift horizontally the experimental point in figure 5-b. However, the operation that allows to jump from figure 5-a to b is not 100% satisfactory as all the corresponding data do not align well on the dashed line. Some other parameters, difficult to take into account (and out of the scope of this study) can also influence band intensities and explain some behaviors. First, the optical constants of the sample (real and imaginary parts of the refractive index, *n* and *k*, respectively) which depends on the chemical composition, drive the light propagation. According to [70], tungsten oxides with O content higher than ≈73 % behaves as a dielectrics whereas for concentrations lower than 73 % the sample behaves as a metal. According to [71], tungsten nitrides are good absorbers but with *k* values which can be ≈ 60% lower than the corresponding value for tungsten. To our best knowledge, nothing is documented on the optical constants of mixed WON samples. However one can imagine that if the k value is lower for $p$WON(20,35) than for the samples aligned on the dashed line: light then probes deeper the sample, thus increasing the intensity of the 800 cm$^{-1}$ band, which could explain part of the deviation from the empirical relation. Another parameter that influences band intensities is surface roughness. For conducting media like W, light penetrates only on a few tens of nm. The scattering process is then very sensitive to the surface structure and an increase of Raman intensities is expected in case of roughness. Here, the roughness is in the range from a few nm to a few tens of nm, as can be seen in figure 2-i and this likely explains the $H_{800}/H_{Si}$ ratio increase (one order of magnitude) from compact $c$WON(20,35) to porous $p$WON(20,35), both having the same chemical composition.

Focusing now on the compact samples, one can see that except for $c$WO(25,2), they all have a lower band intensity than porous samples. For $c$W, tungsten oxide comes from native oxide and the $H_{800}/H_{Si}$ value is representative of WO bonds close to the surface. XPS measurements done after heating the sample to 700°C under UHV conditions allowed us to obtain the atomic composition from relative intensities of W and WO peaks: 21.9% oxygen content in the surface layer, being close to a $W_3O$ stoichiometry [72]. The oxygen content was estimated by performing Angle-resolved XPS experiments. The emission angle at which the electrons were collected was varied from 0 to 30 degrees (with respect to the normal), thereby enabling electron detection from different material depths. Raman data were recorded in air only a couple of hours after the XPS measurement. The Raman spectrum (see figure 3-a) suggests a $WO_6$ octahedral symmetry as discussed above and gives a value of $H_{800}/H_{Si}$ = 0.0027, which may involve a low oxide thickness. Some comments about oxide thickness are given later in the article, in connection with figure 5-c. The $c$WN(0,20) sample contains no oxygen in the bulk. The 800 cm$^{-1}$ band of its Raman spectra, which originates from W-O bonds is then related to the surface. Its intensity is lower than the one of $c$W because not all the tungsten atoms close to the surface are free to bond with surface oxygen but are instead involved in W-N bonds. This is even more pronounced for $c$WON(37,16) where no band at 800 cm$^{-1}$ was detected. According to the empirical law, $H_{800}/H_{Si}$ should be close to 0.01, whereas our analyses bring it lower than 0.0001. This discrepancy could be due to a crystallographic structure that is different from the other samples, leading to no Raman active modes in that spectral region, highlighting the fact that the atomic structure could play a huge role. The $c$WON(20,35) sample is slightly out of the dashed line but the data point can be brought closer to the line ~~on~~ if one does the same treatment as for *p*WON(20,35), replacing O(%) by O(%)×100/(100-N(%)) (i.e. changing from figure 5-a to figure 5-b). However, the cWO(25,2) sample is found one order of magnitude more intense than what suggests the empirical law determined from other samples with similar oxygen content. We believe this is an anomaly caused by the delay between the production, measurement of the oxygen content (one year later) and the Raman measurements of particular samples. Contrary to porous samples, which can develop rapidly an oxide layer with a saturating thickness, more time is needed for bulk materials to reach a saturating oxide layer thickness that acts as a barrier to oxygen diffusion. More detailed kinetic studies are therefore needed to control this phenomenon, which is critical for the future analysis of PFC that will be stored for long period before analysis during a post-mortem analysis phase.

In figure 5-c, we investigate the role of optical constants on the 800 cm$^{-1}$ band intensity, with the help of a simple model developed originally for graphene [73]. The model that gives an enhancement factor has been scaled to reproduce experimental data. It calculates the influence on the Raman band intensities of a layer deposited on top of a semi-infinite media. Multi reflections are then considered in the calculation. Figure 5-c display Raman intensities of WO$_3$ layers, with thicknesses ranging from 20 to 200 nm and grown under controlled conditions (see [35] for details). The measured $H_{800}/H_{Si}$ varies from ≈ 0.1 to ≈ 1, which is one to two decades higher than for the samples discussed previously, which were found in the range ≈ 0.001 to ≈ 0.16. Calculations with several optical constants (n and k) have been done in the range 1-300 nm. Pure oxide is supposed to be n=2.2 and k=0 for $\lambda_L$=514 nm. Introducing some disorder can slightly diminish n, and increase k, the material thus becoming more absorbing. For pure metallic W, the optical constants are n=1.8 and k=4.9 [74, 75]. The first calculation (n=2, k=0-no absorption) reproduces well the experimental data. Extrapolation for intensities in the range close to the one of *c*W leads to a thickness of ≈ 5 nm. Diminishing n and increasing k, which can simulate O and/or N content changes [70, 71], does not change the global shape of the curve below 100 nm. Extrapolation to the *c*W thickness now gives a smaller value of ≈ 3 nm. One conclusion is that such calculations can be used to deduce thicknesses, but it leads to quite large uncertainty as optical constants are not well known. In our case, the conclusion is that for *c*W, native oxide layer thickness is in the range ≈ 3-5 nm. For *c*WO(25,2), if one supposes native oxide behaves as a bulk oxide from an optical property point of view, one obtains a thickness of ≈ 12 nm, which is compatible with a slow growth of native oxide during the one year period discussed above. For porous materials, roughness must be taken into account and thicknesses cannot be retrieved from this simple model.

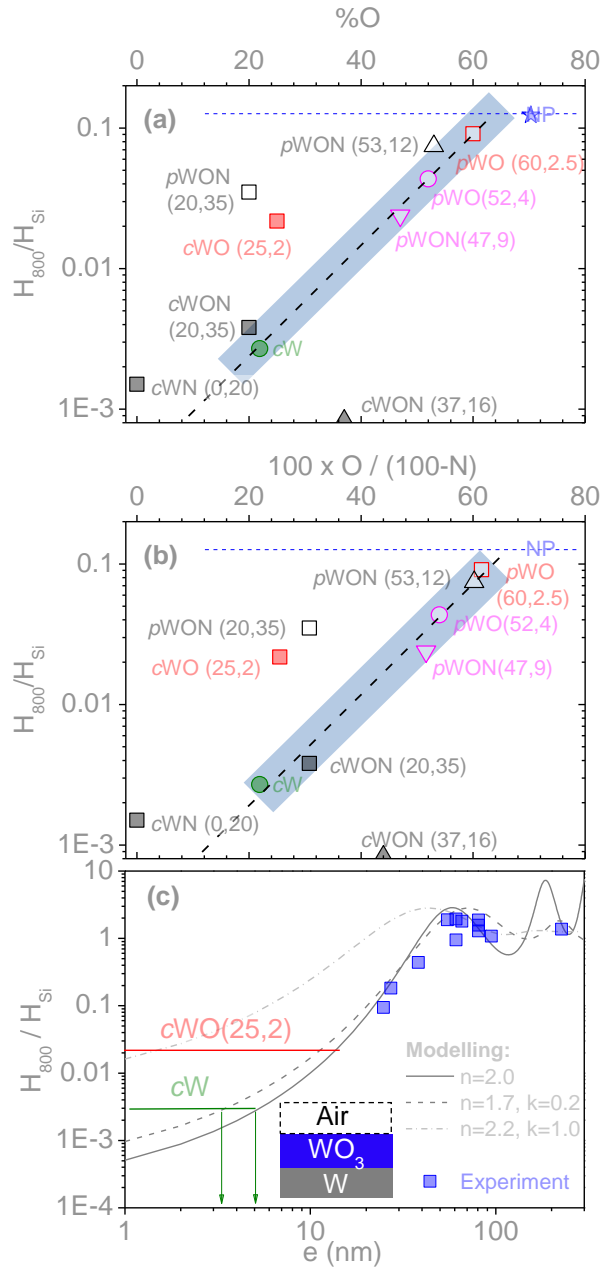

**Figure 5.** *Raman band intensity evolution with respect to the oxygen content. (a) $H_{800}/H_{Si}$ as a function of the O content (the dashed line is a guide for the eyes). (b) Same, but as a function of 100×O/(100-N). In (a) and (b), the dotted blue line represents the nanoparticles value, for comparison. (c) Evolution of $H_{800}/H_{Si}$ for d-WO$_3$ layers as a function of film thickness. The*

*curve represents calculations for different optical indexes. Horizontal lines are experimental values of $H_{800}/H_{Si}$ for cWO(25,2) and cW.*

### 3.3 Laser heating as complementary diagnostic: O/N substitution

As shown in the previous section, compact or porous characteristics can play an important role in the amount of oxides in the material. We address here the question of whether or not oxidation can be induced by local heating due to the Raman laser beam as measurements are performed under ambient atmosphere. We propose below a methodology involving a power variation of the laser intensity up to four orders of magnitude.

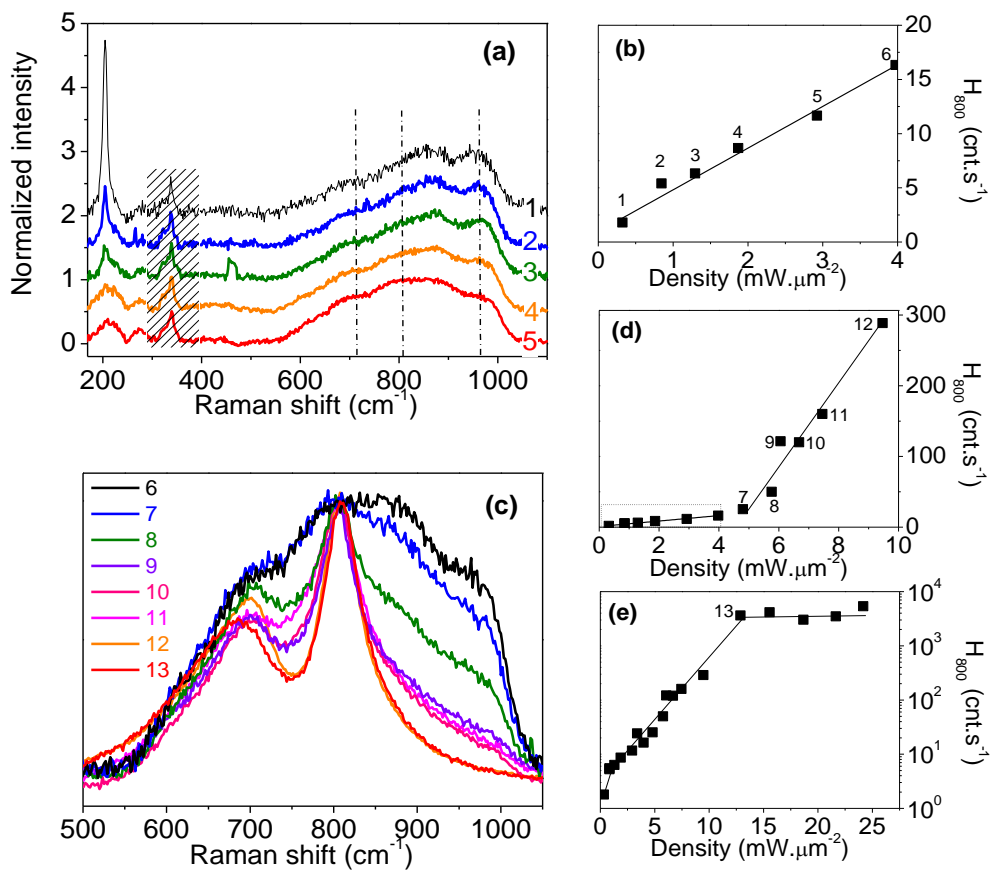

*Figure 6. Influence of laser power density on the pWO(60,2.5) sample. (a) Normalized spectra (vertically shifted for clarity) and (b) 800 cm$^{-1}$ band height ($H_{800}$) as a function of power*

*density (from 0.3 to 4 mW.µm$^{-2}$). (c-e) Normalized spectra and H$_{800}$ as a function of power density for power density higher than 4 mW.µm$^{-2}$. Dashed square of (d) is the range of (b). The density at which the spectra plotted in figure a and c were recorded are given in figure b, d and e.*

Figure 6 shows the evolution of Raman spectra of the *p*WO(60,2.5) sample under different power laser conditions. The spectra in figure 6-a and 6-c are normalized to the height at 800 cm$^{-1}$ and shifted vertically for clarity. The spectra 1 to 5 (6 to 13) of figure 6-a (6-c) correspond to power density varying from 0.3 to 3 (4 to 10) mW.µm$^{-2}$. In figure3-a, in the range 500-1100 cm$^{-1}$, the spectral shape does not evolve, and is composed of three overlapping bands proving the presence of oxides in octahedral forms as discussed in figure 3. For power densities higher than 1.5 mW.µm$^{-2}$, the band at 950 cm$^{-1}$ becomes slightly weaker with respect to the band at 800 cm$^{-1}$, showing that the surface to volume ratio starts to be somewhat modified under the beam. There is a new signature at 205 cm$^{-1}$ and its relative intensity diminishes when increasing the laser power. This signature will be discussed in figure 7. The 800 cm$^{-1}$ band intensity variation with power density is given in figure 6-b, where H$_{800}$ evolves linearly as expected since the Raman signal is proportional to the number of incoming photons. When the power is increased beyond 4 mW.µm$^{-2}$, changes in the 500-1100 cm$^{-1}$ region become important (figure 6-c): the 950 cm$^{-1}$ band significantly diminishes compared to the other bands, and the overlap between the bands close to 700 and 800 cm$^{-1}$ decreases, behaving more and more as an organized phase, resembling WO$_3$ (see figure 1). In figure 6-d, the H$_{800}$ versus power density curve increases its slope drastically for power density higher than a threshold of 4 mW.µm$^{-2}$. Figure 6-e, in semilogarithmic plot on the vertical axis, shows the existence of a saturation regime reached for this sample close to 12 mW.µm$^{-2}$. These observations indicate that a well-organized WO$_3$ layer is formed under the laser beam and therefore that the temperature is sufficiently high to activate oxidation mechanisms in the presence of ≈ 0.2 bar of O$_2$ and traces of water (a few hundreds of kelvins according to [55] and references therein. See also the discussion on the last figure of this paper). This is possible only if heat cannot be evacuated by the sample and this principally occurs when the morphology is not optimal for heat dissipation. Then, reaching the laser power density threshold allows atoms to reorganize and change structure. To prove this qualitative argument, we show in figure 7 the same kind of graph for the three studied morphologies: compact (figure 7-a), porous (figure 7-b) and nanoparticles (figure 7-c). Compact sample spectra do not evolve on the whole range of power density tested here (up to 24.15 mW.µm$^{-2}$), whereas for the porous sample and nanoparticles, the shapes in the 500-1100 cm$^{-1}$ region (which are similar but with a higher 950 cm$^{-1}$ band for nanoparticles) change from disordered to a more organized oxide at 4 and 1 mW.µm$^{-2}$, respectively. The power at which new oxide is created is higher for the porous sample than for nanoparticles, indicating that the porous film can dissipate heat more efficiently than nanoparticles (which are agglomerated). The Raman spectral shape corresponding to the more organized samples differs in the two cases: it is similar to what is expected for well-organized $WO_3$ in the porous sample whereas for nanoparticles the main signal comes from the band at 950 cm$^{-1}$. This reveals a phase which has a large surface / volume ratio (with no coalescence of nanoparticles when the power increases). The band observed at 205 cm$^{-1}$ has already been observed previously and tentatively attributed to the presence of small domains and stoichiometric defects [56]. It is not often reported in the literature and its interpretation is still under debate. We show here that it disappears when the applied laser power density is too high. Note that this band is not observed for compact samples but is observed, although down-shifted by 65 cm$^{-1}$, for nanoparticles: this indicates that it is related to surface vibrators, confirming the previous interpretation involving small domains and/or defects.

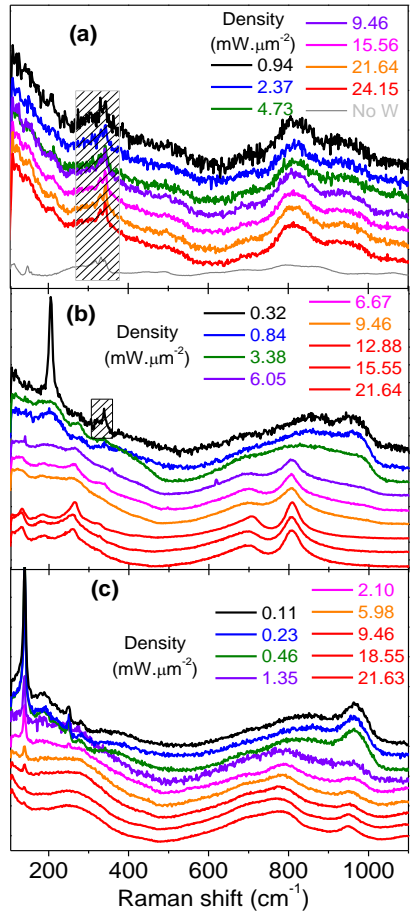

*Figure 7. Influence of morphology on sample evolution under laser irradiation. (a) Compact WO, cWO(25, 2). (b) Porous WO, pWO(60, 2.5). (c) Nanoparticles.*

To systematically investigate the power density at which the laser beam creates oxide, the intensity of the oxide band at 800 cm$^{-1}$ is compared to that of silicon at 520 cm$^{-1}$, recorded in the same experimental conditions. If no evolution with power density is expected, then as the band intensities of oxide and silicon are both proportional to the number of incoming photons (and thus to the power density of the used laser beam), $H_{800}/H_{Si}$ should remain constant. On the other hand, if oxides are created, as shown in figure 6, $H_{800}/H_{Si}$ will evolve with the laser power density. In figure 8, we plot this ratio for some compact and porous samples as a function of the laser power density, for samples without nitrogen in figure 8-a, and for samples with both oxygen and nitrogen in figure 8-b. Figure 8-a shows that, as expected for compact samples, for *c*W and *c*WO (25,2), the ratio remains constant within the entire range of investigated power densities (dispersion around this value comes from the size of the spot which could change a little from one measurement to another), meaning the oxides probed by Raman spectroscopy are not created but are inherent to the pristine samples. On the other hand for porous samples and nanoparticles, there is a clear threshold (close to 4 for *p*WO(52,4), *p*WO(60,2.5) and 1 mW.µm$^{-2}$ for NP) after which the ratio starts to increase drastically, up to $H_{800}/H_{Si} \approx 5$. According to the model presented in figure 5-c, this value corresponds to a thickness of about 50 nm. This is of course a very approximate thickness estimation as the model is based on a simple slab geometry. In reality, samples are porous and the oxide created by the laser beam is almost circular in the x, y plane with a radius close to that of the laser spot, meaning the oxide created is surrounded by porous tungsten containing various amounts of O, depending on the sample. On the contrary, figure 8-b shows that, for compact samples, except in the case of *c*WN(0,20) which behaves similarly to *c*W, *c*WON(20,35) and *c*WON(37,16) starts to evolve at a threshold of ≈ 10 and 8 mW.µm$^{-2}$, respectively (with no detectable oxide for the latter below this threshold). These results show that the presence of N likely favor the creation of oxides in compact layers. On the other hand, the three porous samples (*p*WON(53,12), *p*WON(47,9) and *p*WON(20,35)) behave more or less similarly to each other and to the one without N: the $H_{800}/H_{Si}$ ratio is constant until a threshold is reached, and then increases. The thresholds are ranging from 2 to 5 mW.µm$^{-2}$ and, contrary to compact samples, there is no evidence on the role of nitrogen.

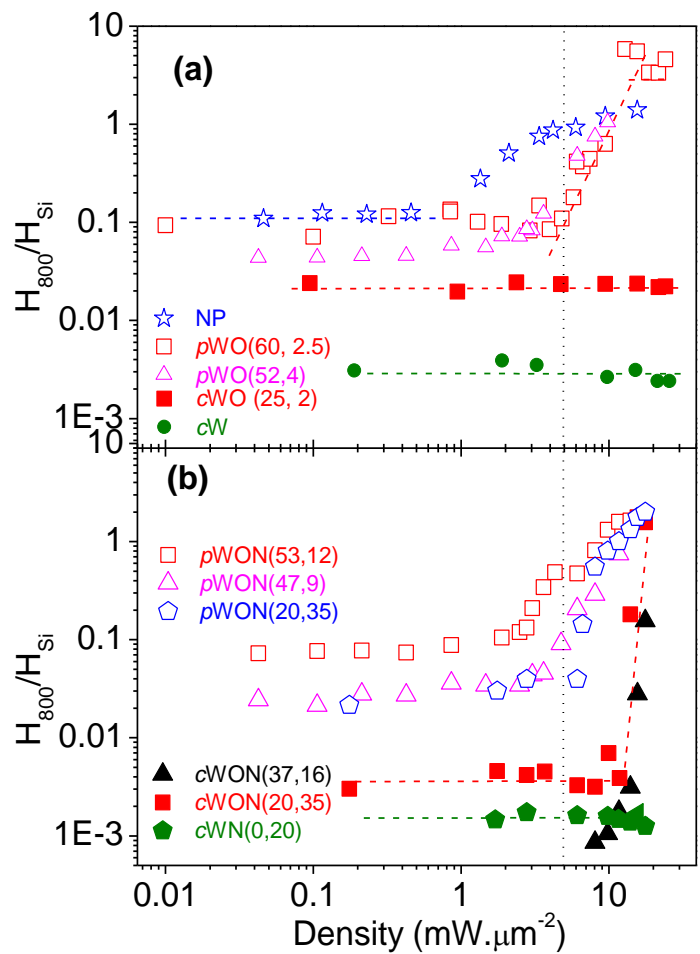

**Figure 8.** *Oxide Raman band intensity evolution as a function of laser power density. The $H_{800}/H_{Si}$ ratio for (a) cW, nanoparticles, and PLD samples without nitrogen and (b) PLD samples containing nitrogen.*

**3.4 Qualitative information on the sample temperature during oxide formation.**

When increasing the power density of the laser beam under atmospheric conditions, we have seen previously that sample chemistry can evolve, due to the local heating creating bulk oxides. However, since the laser beam which locally heats the sample is also the beam that probes the sample on the exact same spot, the retrieved spectrum is recorded at a sample temperature, which can in principle be higher than room temperature. Figure 9-a shows two spectra taken from the *c*WON(37,16) sample: one recorded when the power density was 17 mW.µm$^{-2}$, the other recorded after this measurement, at a lower power density of 1 mW.µm$^{-2}$, for which the sample is expected to be close to room temperature. The two spectra are different: at low laser power densities, the 800 cm$^{-1}$ band slightly upshifts (4 cm$^{-1}$) and narrows while the band at 700 cm$^{-1}$ gets more intense and better defined. To obtain hints on the sample temperature due to laser heating, we studied a reference WO$_3$ sample, heated into a furnace (1 bar argon atmosphere, temperatures varied from 20 to 1050°C) equipped with a window that allowed *in-situ* Raman spectroscopy to be performed. Figure 9-b shows the evolution of the Raman spectrum of this WO$_3$ sample as a function of temperatures. The laser power was kept low enough to avoid local heating. When the temperature increases, the 800 cm$^{-1}$ band downshifts and broadens and the band at 700 cm$^{-1}$ starts to disappear. The effect is reversible (not shown here). Note that WO$_3$ is polymorphic and present a series of transition between 17°C and 740 °C, from $\delta$-WO$_3$ to $\gamma$-WO$_3$ to $\beta$-WO$_3$ to $\alpha$-WO$_3$ (see section 2.3). The 1050°C spectrum then corresponds to a spectrum of the $\alpha$−WO$_3$ phase. This spectrum being similar to that observed for the sample heated at 17 mW.µm$^{-2}$ and recorded at 17 mW.µm$^{-2}$ (figure 9-a), one can conclude that it corresponds to $\alpha$−WO$_3$, whereas the subsequent 1 mW.µm$^{-2}$ corresponds to $\gamma$−WO$_3$ (with defects, as widths are larger than the reference well crystallized sample). However, one cannot conclude exactly on the value of the reached temperature due to laser heating because Raman spectra of WO$_3$ are sensitive to the crystal order and to the presence of defects (here for example, the two spectra at room temperature of figure 9-a and 9-b are not strictly similar). However, one can conclude the sample reached an elevated temperature.

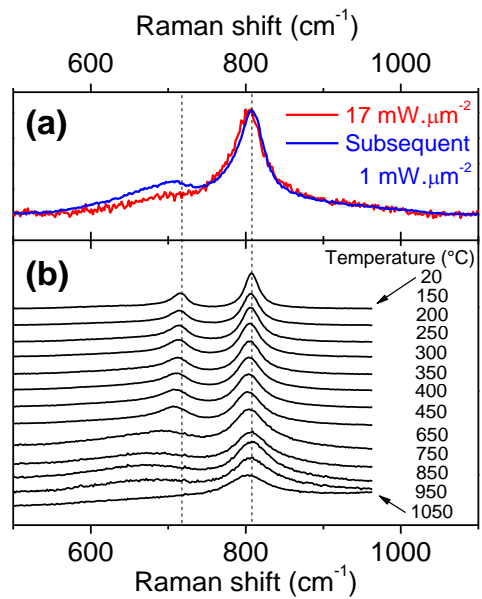

*Figure 9. Crystalline phase cWON(37,16) heated at 17 mW µm². (a) Spectra recorded at 17 mW.µm⁻² and at 1 mW.µm⁻² after the 17 mW µm² irradiation (spectra have been normalized for clarity) (b) Spectra of a reference $WO_3$ sample recorded in a heating cell (see details in the text).*

## 4. Discussion and concluding remarks

We have investigated the sensitivity of Raman spectroscopy to tungsten oxide and nitride that could be formed in compact, porous and nanoparticle forms. In the perspective of a tokamak post-mortem campaign, this technique has the advantage to give rapidly a large amount of data based on semi-qualitative spectroscopic observations. This could be used to select samples that need to be quantitatively analyzed with more time-consuming techniques, like TOF-ERDA or electron microscopies. The main spectroscopic findings observed on relevant laboratory samples are:

(1) Raman spectroscopy is sensitive enough to detect surface native oxides, and the optimal experimental conditions to perform such measurements were derived
(2) More tungsten oxides are detected for porous materials and nanoparticles than for compact ones due to the large roughness of the former class of materials
(3) Raman band intensity varies and increases qualitatively with the O content
(4) The crystallographic structure has an influence on band intensities
(5) Long-time storing of a sample between its production and analysis can affect the results because of the growth native oxide in the meantime
(6) Porous materials are less thermally stable than compact materials
(7) The relative thermal stability of compact materials is deteriorated by an increasing nitrogen content
(8) Oxides and nitrides can be detected and distinguished (directly by the presence of additional broad bands in the range 300-600 cm⁻¹, and indirectly by the W PDOS activation close to 200 cm$^{-1}$, for nitrides), even if sensitivity to N is lower than that of O

(9) An oxygen/nitrogen substitution occurs when the temperature is high, and oxide is created under the laser beam

We believe this work will also be useful for Laser cleaning activities [76], mirror activities [77-80], tungsten oxide engineering activities [54, 81-84], N/O substitution in W, and characterization of new tungsten nitride phases [85], involved for example in water splitting [86]. Two natural continuations of this work could be: studying the kinetics of growth after exposure to air, and detection/quantification of hydrogen isotopes introduced in these materials.

Note that to obtain more quantitative results, details of light propagation inside these complex materials should be considered. The stoichiometry as well as the roughness and the porosity can change optical properties and thus Raman band intensities. For example, oxides with a O/W atomic ratio of 2.7 are at the edge of a metallic/insulator transition [70], thus with totally different optical properties.

The phenomenological points evidenced in this paper, if Raman spectroscopy is used as the first post-mortem technique could be of importance for systematic measurement done in the framework of post-mortem campaigns as these measurements could be done rapidly on many samples. Moreover, statistics could be achieved in a reasonable time scale, even on a single sample, due to the lateral micrometric resolution

**Acknowledgements**

This work has been carried out within the framework of the EUROfusion Consortium and has received funding from the Euratom research and training programme 2014-2018 and 2019-2020 under grant agreement No 633053. The views and opinions expressed herein do not necessarily reflect those of the European Commission. Work performed under EUROfusion WP PFC.

M. M. has received funding from the Excellence Initiative of Aix-Marseille University – A*Midex, a French "Investissements d'Avenir" programme as well as from the ANR under grant ANR-18-CE05-12